\documentclass[11pt]{article}
\usepackage[utf8]{inputenc}
\usepackage[T1]{fontenc}
\usepackage[normalem]{ulem}
\pdfoutput = 1

\usepackage{amsmath}
\usepackage{amssymb}
\usepackage{bbm}
\usepackage{bbold}
\usepackage{braket}
\usepackage{caption}
\usepackage{color}
    \definecolor{darkgreen}{rgb}{0,0.5,0}
    \definecolor{darkred}{rgb}{0.5,0,0}
    \definecolor{darkblue}{rgb}{0,0,0.6}
    \definecolor{purple}{rgb}{0.4,.2,0.7}
\usepackage{comment}
\usepackage{csquotes}
\usepackage[mathcal]{eucal}
\usepackage{graphicx}
\usepackage[hyperfootnotes = true, colorlinks = true, linkcolor = blue, citecolor = blue,urlcolor = blue]{hyperref}
\usepackage{mathabx}
\usepackage{mathtools}
\usepackage{pdfsync}
\usepackage{slashed}
\usepackage{url}
\usepackage{upgreek}

\usepackage[margin = 2.2cm]{geometry}
\usepackage[ragged]{footmisc}
\renewcommand{\d}{\mathrm{d}}
\renewcommand{\i}{\mathrm{i}}

\begin{document}

\thispagestyle{empty}
\begin{center}
    ~\vspace{5mm}
    
    {\Large \bf 

        A Brief Note on Complex AdS-Schwarzschild Black Holes
    }
    
    \vspace{0.4in}
    
    {\bf Raghu Mahajan$^1$ and Kaustubh Singhi$^{1}$}

    \vspace{0.4in}

    $^1$ International Centre for Theoretical Sciences,  Shivakote Village,  Hesaraghatta Hobli,  \\
    Bengaluru 560089,  Karnataka,  India
    
    {\tt raghu.mahajan@icts.res.in, kaustubh.singhi@icts.res.in}
\end{center}

\vspace{0.4in}

\begin{abstract}
In the context of thermodynamics of asymptotically anti-de Sitter spaces, it is often stated that at very low temperatures, there is only one saddle point available---namely, thermal AdS---and hence this sole saddle dictates the low-temperature behavior.
However, AdS-Schwarzschild black holes continue to exist at low temperatures as complex saddle points.
We point out that the real part of the on-shell action of these complex black holes is \emph{smaller} than that of thermal AdS at the lowest temperatures, in AdS$_5$ and higher dimensions.
So, na\"{\i}vely, they should be the ``dominant'' saddles.
This raises a puzzle: if these complex black holes were indeed the relevant saddle points, the physics of the bulk and that of the dual gauge theory would completely disagree at low temperatures.
Using a mini-superspace approximation and contour arguments, we argue that these complex black holes do not actually contribute to the gravitational path integral, regardless of the value of their on-shell action.
So the standard conclusion that thermal AdS is the correct saddle at the lowest temperatures continues to hold.
We also comment on two related matters: whether the Kontsevich--Segal criterion is useful in this setting, and whether the unstable small black hole contributes to the path integral in the high-temperature phase.

\end{abstract}

\pagebreak


A basic observable that encodes the thermodynamics of a CFT$_d$ is its partition function on $S^{d-1}\times S^1$.
Setting the radius of $S^{d-1}$ to one and the circumference of the $S^1$ to $\beta$, we denote this quantity by $Z(\beta)$. 
For theories that admit a gravity dual \cite{Maldacena:1997re, Gubser:1998bc, Witten:1998qj}, $Z(\beta)$ can be computed using the gravitational path integral with boundary conditions that fix the metric to be that of $S^{d-1}\times S^1$ at the boundary of the $(d+1)$-dimensional spacetime.
It was realized soon after the discovery of AdS/CFT that the Hawking-Page transition in asymptotically AdS spacetimes \cite{Hawking:1982dh} corresponds to the confinement-deconfinement transition in the dual gauge theory \cite{Witten:1998zw}.
This correspondence between the phase transition in gravity and the phase transition in the dual gauge theory is one of the beautiful cornerstone results in the AdS/CFT correspondence \cite{Aharony:2003sx}.

Let us quickly summarize the known facts, restricting our attention to $d \geq 3$.
One relevant bulk saddle, which exists for all $\beta$, is the thermal AdS$_{d+1}$ space, with the line element
\begin{align}
    \d s^2 &= \left( 1+ \frac{r^2}{L^2} \right) \d \tau^2 
    + \frac{\d r^2}{1 + r^2/L^2} + r^2 \d \Omega_{d-1}^2 \,, \label{thermal-ads-metric} \\
    \tau &\equiv \tau + \beta \, .
\end{align}
The partition function $Z(\beta)$, and hence the entropy, can be computed using a saddle point approximation around this saddle point. 
One finds that the entropy does not have an $O(\frac{1}{G})$ scaling, which is fundamentally due to the fact that (\ref{thermal-ads-metric}) does not have a horizon.

Other asymptotically AdS saddles with $S^{d-1} \times S^{1}$ boundary conditions exist, namely, the AdS Schwarzschild black holes, which have the line element
\begin{align}
    \d s^2 &= f(r) \d \tau^2 + \frac{\d r^2}{f(r)} + r^2 \d \Omega_{d-1}^2\,, \quad \text{where} \label{ads-sch-metric} \\
    f(r) &= 1 + \frac{r^2}{L^2} - \frac{r_+^{d-2} (1 + r_+^2/L^2)}{r^{d-2}}\, ,\quad \text{and} \label{def-fr}\\
    \beta &= \frac{4\pi}{f'(r_+)} = \frac{4\pi L}{d \, \frac{r_+}{L} + (d-2)\, \frac{L}{r_+}}\, . \label{beta-rplus-relation}
\end{align}
The horizon is located at $r=r_+$, and the relationship (\ref{beta-rplus-relation}) between $\beta$ and $r_+$ is fixed using the requirement that the Euclidean metric (\ref{ads-sch-metric}) be smooth at the horizon \cite{Gibbons:1976ue}.
Note that the local form of the line element (\ref{ads-sch-metric}) reduces to that of thermal AdS (\ref{thermal-ads-metric}) in the limit $r_+ \to 0$. 
However, we must remember that $\beta$ in the thermal AdS saddle is a free parameter.

Following the logic of Gibbons and Hawking \cite{Gibbons:1976ue}, 
and after including the appropriate boundary counterterms \cite{Skenderis:2002wp},
we can compute the on-shell action of the black hole saddles. 
Setting $L=1$ and specializing to $d=4$ for concreteness,\footnote{For asymptotically AdS$_5$ spaces, the action is 
$I_\text{Euc} = - \frac{1}{16\pi G} \int \d^5 x \, \sqrt{g}\, (R + 12) - \frac{1}{8\pi G} 
\int d^4 x \, \sqrt{\gamma} \left( K - 3- \frac{1}{4} \, R^{(4)}\right)$.} one finds
\begin{align}
    \log Z(\beta) \approx \frac{1}{16\pi G} \cdot 2\pi^2 \cdot \beta \left( 
    -\frac{3}{4} + r_+(\beta)^4 - r_+(\beta)^2
    \right)\, . \label{log-Z-bh}
\end{align}
For the record, we note that the entropy and energy computed using (\ref{log-Z-bh}) are
\begin{align}
    E &= \frac{3\pi}{32G} \left( 1 + 4 r_+^2 + 4 r_+^4 \right) \,, 
    \label{e-rplus-relation} \\
    S &= \frac{1}{4G} \cdot 2\pi^2\cdot r_+^3 \, .
    \label{s-rplus-relation}
\end{align}

\begin{figure}
    \centering
    \includegraphics[width=0.45\textwidth]{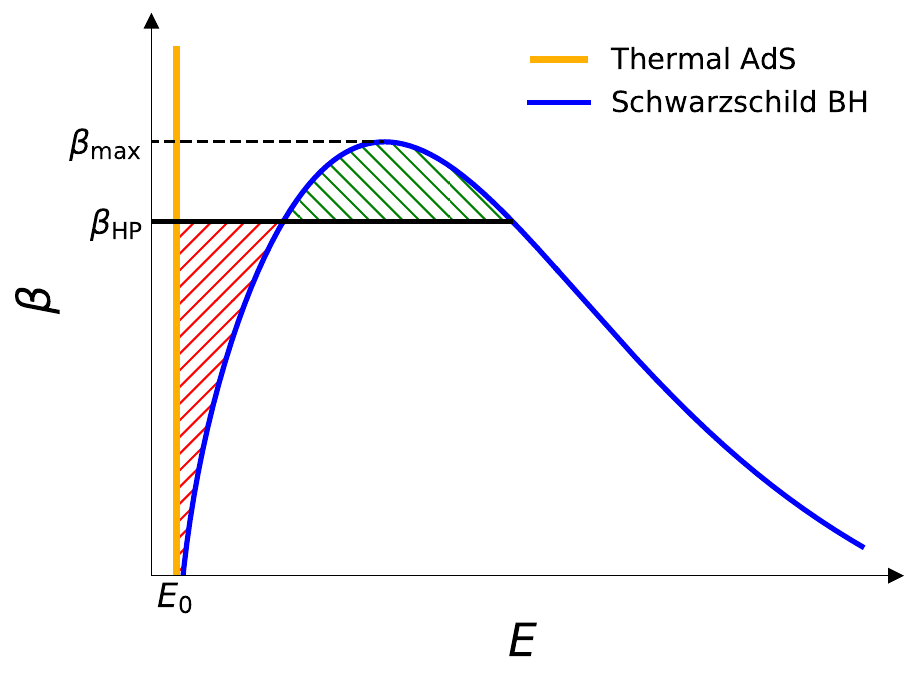}
    \caption{A plot showing the thermal AdS and the Schwarzschild black hole saddles in the $\beta-E$ plane. The quantity $E_0$ can be interpreted as the Casimir energy of the field theory on $S^{d-1}$. 
    The Hawking-Page transition happens at $\beta_\text{HP}$. 
    The branch of the blue curve with positive slope is the small black hole, and the branch with negative slope is the big black hole.
    The red and green shaded areas are equal, a version of Maxwell's equal area construction for first-order phase transitions.}
    \label{fig:beta-vs-E}
\end{figure}

It follows from (\ref{beta-rplus-relation}) that as $r_+$ goes from zero to infinity, $\beta$ goes from zero to a maximum value, $\beta_\text{max}$, and then goes back down to zero \cite{Hawking:1982dh}.
This also implies that for $\beta < \beta_\text{max}$, there are two values of $r_+$; they are usually called the ``small'' and the ``big'' black hole, with the big black hole having the larger value of the horizon radius.
Since we are computing $Z(\beta)$, it is $\beta$ which should be thought of as the independent variable. 
So we have three candidate gravitational saddles for $\beta < \beta_\text{max}$.
One can proceed to compare the free energies of the various saddles,\footnote{The free energy is defined as $-\log Z$ in our conventions.} and find which solution is ``dominant'', with lower free energy.
One finds that there exists a $\beta_\text{HP}$ with $0 < \beta_\text{HP} < \beta_\text{max}$ such that, for $\beta > \beta_\text{HP}$, the thermal AdS saddle is dominant, and for $\beta < \beta_\text{HP}$, the big black hole is dominant.
Finally, note from (\ref{e-rplus-relation}) that the energy $E$ is a monotonic function of $r_+$.
We can concisely summarize the above facts using an $E-\beta$ plot, shown in figure \ref{fig:beta-vs-E}.

\begin{figure}
    \centering
    \includegraphics[width=0.35\linewidth]{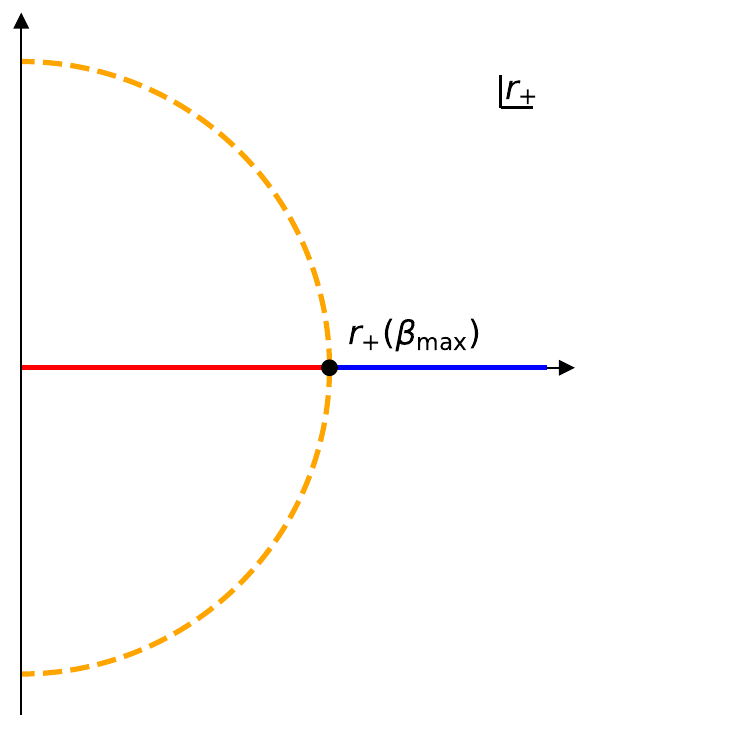}
    \caption{The trajectory $r_+(\beta)$ in the complex $r_+$ plane upon varying $\beta$.
    The small black hole saddle (red) and the big black hole saddle (blue), which exist for $\beta < \beta_\text{max}$ merge at $\beta = \beta_\text{max}$. 
    For $\beta > \beta_\text{max}$, the two possible values of $r_+(\beta)$ are complex. These complex AdS Schwarzschild black holes are shown in orange.}
    \label{fig:rplus-plane}
\end{figure}

Our main concern in this note is the simple fact that the relation (\ref{beta-rplus-relation}) between $\beta$ and $r_+$ continues to hold even when $\beta > \beta_\text{max}$, albeit with the two possible values of $r_+$ now being complex;
see figure \ref{fig:rplus-plane}.\footnote{See also \cite{DiTucci:2020weq}, where these complex black holes were discussed in the AdS$_4$ case with the boundary at a finite cutoff.}
Should we consider the corresponding geometries, specified by (\ref{ads-sch-metric}), (\ref{def-fr}) and (\ref{beta-rplus-relation}), as candidate saddle points in the canonical ensemble or not?
Do they contribute to $Z(\beta)$?
These geometries with complex values of $r_+$ have a complex metric, but that does not a priori mean that we should discard them.
Indeed, by now it is well appreciated that complex geometries give rise to interesting physical effects 
in cosmology \cite{Maldacena:2024uhs, Ivo:2024ill, Feldbrugge:2017kzv, Jonas:2022uqb, Hertog:2011ky, Chen:2020tes},
in the gravitational description of quantum chaos \cite{Saad:2018bqo, Saad:2019pqd, Stanford:2020wkf, Chen:2023hra}, 
and in real-time AdS holography \cite{Skenderis:2008dh, Glorioso:2018mmw, Jana:2020vyx, Loganayagam:2024mnj}.

When $\beta \to \infty$, the two values of $r_+$ become almost purely imaginary, and so the combination $r_+^4 - r_+^2$ appearing in (\ref{log-Z-bh}) will be positive.
This means that these complex saddles are more dominant compared to thermal AdS, for which $\log Z(\beta) = - \frac{3}{4} \frac{2\pi^2 \beta}{16\pi G}$. 
So na\"{i}vely we would conclude that the partition function at low temperatures (large $\beta$) should be controlled by these complex black holes and not by thermal AdS.
Needless to say, this would be disastrous for the AdS/CFT correspondence \cite{Witten:1998zw, Aharony:2003sx}, since thermal AdS very well captures the low-temperature phase of the dual gauge theory.\footnote{It would perhaps be interesting to find the analogs of these complex black holes in the holonomy-eigenvalue distributions that are used for analyzing the phase structure of weakly-coupled gauge theories \cite{Aharony:2003sx}. In particular, do saddle points other than the uniform distribution on the circle exist at low temperatures?}

In the absence of a more refined argument, it is hard to decide this issue one way or another.
An ad hoc way would be to invoke the Kontsevich--Segal criterion \cite{Kontsevich:2021dmb, Witten:2021nzp} which disallows certain complex metrics.
The black holes which have $\text{arg}(r_+)$ larger than an order-one critical value will be excluded by this criterion.

However, we can address this issue in a more systematic way by writing a mini-superspace approximation for the full gravitational path integral.
The main idea is to note that since we are fixing $\beta$ via the boundary condition, the energy is one of the integration variables in the path integral.
The energy is monotonically related to $r_+$, so we can instead think of $r_+$ as an integration variable.\footnote{We are only interested in the classical limit, so we will not worry about any Jacobian factors that will affect the result only at one-loop order.}
We approximate the full path integral as just a one-dimensional integral over $r_+$.
In this way of thinking, the relation (\ref{beta-rplus-relation}) between $r_+$ and $\beta$ arises as a saddle point equation in this integral.
Versions of this particular mini-superspace approximation have appeared earlier in the literature \cite{Marolf:2018ldl, Dong:2018seb, Akers:2018fow, Marolf:2022ybi}.
So we study the following integral:
\begin{align}
    Z(\beta) = \int_0^{\infty} \d r_+ \exp \left( 
    \frac{2\pi^2 r_+^3}{4G} 
    - \beta \, \frac{3\pi}{32G} \left( 1 + 4 r_+^2 + 4 r_+^4 \right) 
    \right)\,.
    \label{zbeta-minisuperspace}
\end{align}
The argument of the exponential is simply $S(r_+) - \beta E(r_+)$, with $S(r_+)$ and $E(r_+)$ being the same as in (\ref{e-rplus-relation}) and (\ref{s-rplus-relation}).
This can be derived by computing the Einstein action (with the boundary terms at infinity) on the line element (\ref{ads-sch-metric}) for general $r_+$ and $\beta$. 
In particular, there is a nonzero contribution $\frac{2\pi^2 r_+^3}{4G} \left( 1 - \frac{f'(r_+)}{4\pi} \,\beta \right)$ from the delta function in the Ricci scalar at the conical singularity $r=r_+$.
An essential point about (\ref{zbeta-minisuperspace}) is that we are making the reasonable assumption that the contour of integration in the $r_+$ plane is along the real axis, corresponding to real values of the energy.

\begin{figure}
    \centering
    \includegraphics[width=0.5\textwidth]{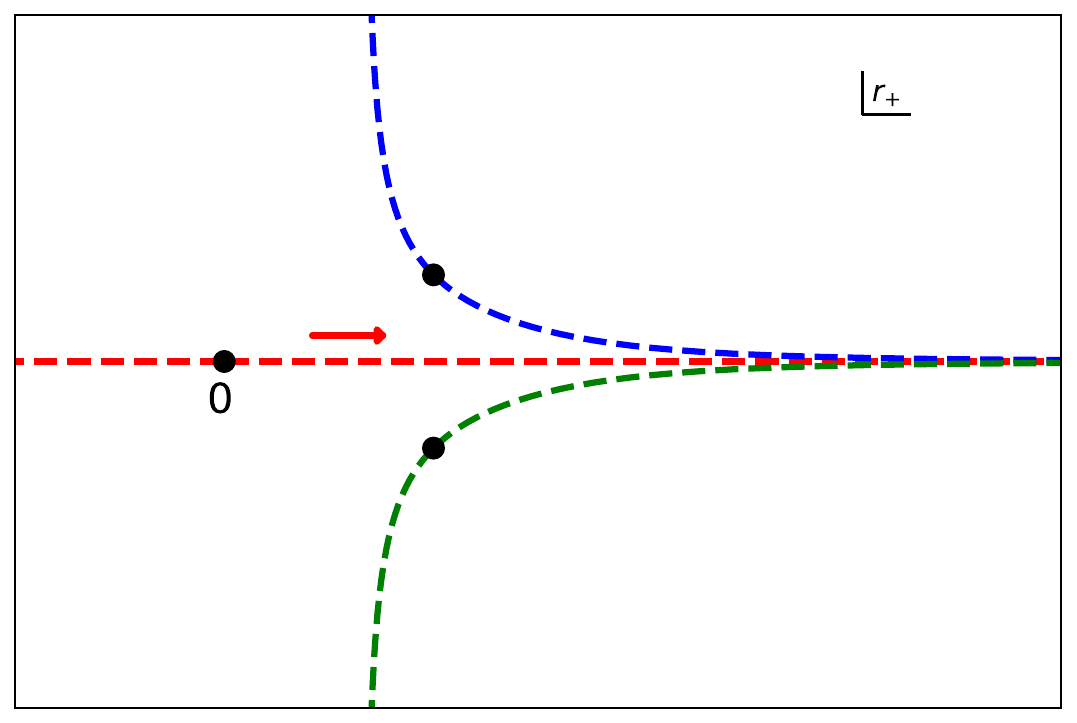}
    \caption{The steepest descent contours of the two complex black hole saddles, which exist for $\beta > \beta_\text{max}$, are shown in blue and green. 
    The red contour is the steepest descent contour of the thermal AdS saddle.
    The saddle locations are shown by thick black dots.
    The defining contour is the positive real line, and thus is (half of) the steepest descent contour of the thermal AdS saddle.
    This shows that the complex black holes do not contribute to $Z(\beta)$.}
    \label{fig:complex_bh_steepest_descent}
\end{figure}

The ``action'' in (\ref{zbeta-minisuperspace}) is a polynomial in $r_+$ containing quadratic, cubic and quartic terms.
In particular, one of the saddle points is always $r_+ = 0$ and its steepest descent contour is always along the real-$r_+$ axis, due to the negative coefficient of the $r_+^2$ term in (\ref{zbeta-minisuperspace}).
Indeed, this $r_+ = 0$ saddle is nothing but the thermal AdS saddle.
It is a general fact that if the defining contour of an integral is the steepest descent contour of one particular saddle point, and if this contour does not pass through any other saddle points, then only this particular saddle point contributes to the integral.\footnote{See, for example, the discussion around Eq.(3.22) of \cite{Witten:2010cx}. In our case the defining contour is half of the steepest descent contour of the thermal AdS saddle, but the result still applies.}
Thus, we have shown that for $\beta > \beta_\text{max}$, the complex Schwarzschild black holes, which are complex saddle points not lying on the integration contour, do not contribute to $Z(\beta)$, since the integration contour is the steepest descent contour of the thermal AdS saddle point.
For slightly more detail, we refer to figure \ref{fig:complex_bh_steepest_descent}.

Note that for $\beta > \beta_\text{max}$, but only slightly, the complex black holes will remain allowed by the Kontsevich--Segal criterion, and will also be ``subdominant'' compared to the thermal AdS saddle.
However, we have presented a more refined argument that shows that these complex black holes do not contribute to $Z(\beta)$, no matter how small $\text{Im}(r_+)$ is.\footnote{There are also examples where geometries that are na\"{i}vely excluded by the Kontsevich--Segal criterion are essential to reproduce the results from a dual description \cite{Mahajan:2021nsd, Maldacena:2019cbz}. Specifically, it is a two-sphere geometry with a $(-,-)$ signature which reproduces (the cutoff-independent part of) the sphere partition function in minimal string theory.}

Let us also make a remark about Ref. \cite{DiTucci:2020weq} which studied these complex black holes in the AdS$_4$ case.
As observed in \cite{DiTucci:2020weq}, the real part of the on-shell action of these complex black holes in the AdS$_4$ case is never smaller than that of thermal AdS$_4$. 
The conclusion of \cite{DiTucci:2020weq} was that the complex black holes contribute to the thermodynamics, albeit in a suppressed way compared to thermal AdS$_4$. 
However, our argument shows that the complex black holes do not contribute even in a suppressed way, in any dimension.

\begin{figure}
    \centering
    \includegraphics[width=0.49\textwidth]{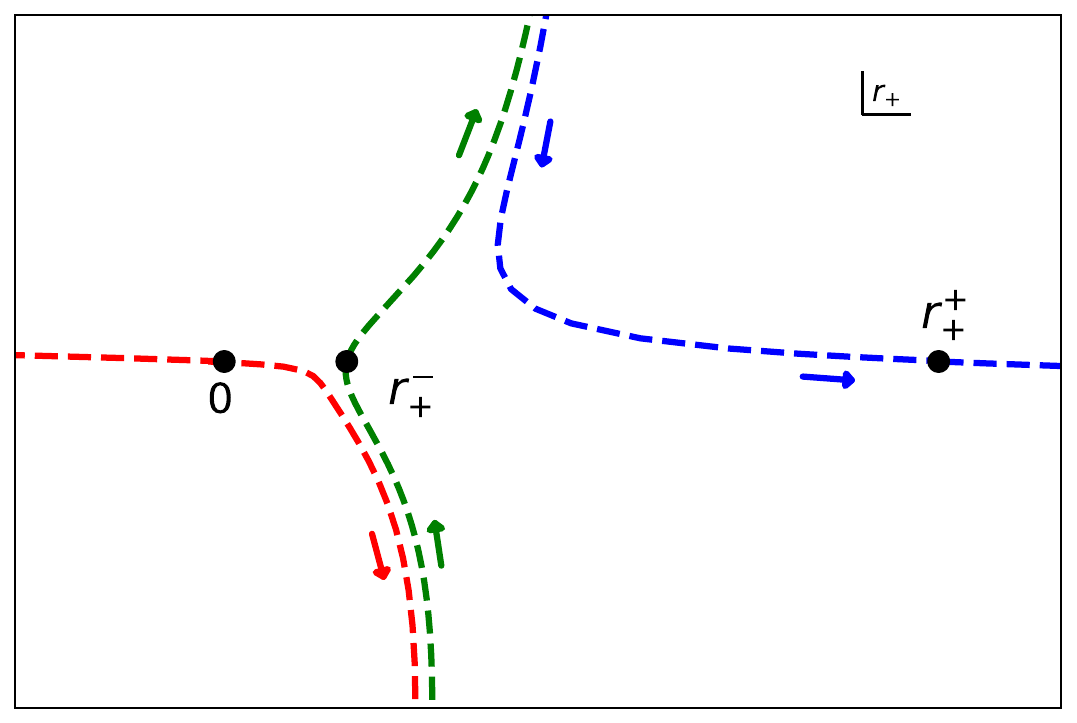}
    \includegraphics[width=0.49\textwidth]{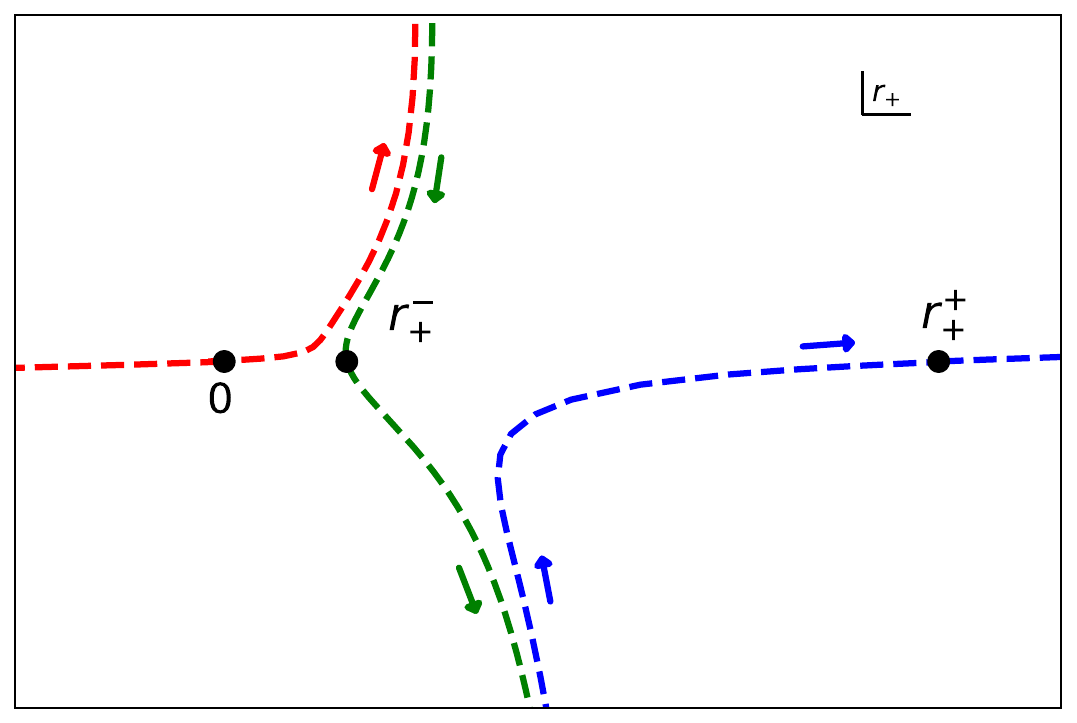}
    \caption{The steepest contours through the three saddle points with $G$ having negative imaginary part (left) or positive imaginary part (right), for $\beta < \beta_\text{max}$. 
    The saddle at zero is thermal AdS, the one labeled $r_+^-$ is the small black hole saddle and the one labeled $r_+^+$ is the big black hole.
    We can approach the limit of real $G$ with either sign of the imaginary part; the answer for the integral will be the same.}
    \label{fig:complexG}
\end{figure}
Finally, we can ask another question: For $\beta < \beta_\text{max}$, does the small black hole saddle contribute to the path integral?
This is a tricky question and, in some sense, needs a bit of refinement.
In more detail, for $\beta < \beta_\text{max}$, the imaginary part of the on-shell action of all three saddles is zero, and thus we are precisely on a Stokes ray in the space of parameters ($\beta$ and $G$).
There is a local maximum of the action sandwiched between two minima, all on the integration contour.
The cleanest way to proceed is to complexify a parameter and break the degeneracy between the imaginary parts of the on-shell actions on the various saddle points.
In appendix F of \cite{Maldacena:2019cbz}, this question was studied for the case of four-bulk dimensions by complexifying $\beta$.
We will present an analysis by complexifying the Newton constant $G$, whose smallness justifies the saddle-point approximation.\footnote{See \cite{Maldacena:2024spf} for another recent example where complexifying $G$ is necessary.
See also appendix A of \cite{Lee:2024hef} for a closely related recent discussion of a similar model integral with the action having a double-well structure.}

So, let us give a small imaginary part to $G$, with $\text{Re}(G) > 0$.
The defining contour is homologically equivalent to, and so can be deformed to, a sum of \text{all three} steepest descent contours.
The precise structure is highly sensitive to whether the imaginary part is positive or negative, see figure \ref{fig:complexG}.

Despite the apparent difference, it is easy to see that the answer to the integral will be the same whether we approach the limit of real $G$ with a small positive or a small negative imaginary part.
This had better be the case since the integral (\ref{zbeta-minisuperspace}) is completely convergent and well-defined for real positive $G$.
Indeed, the difference between the contours shown in the left and right panels of figure \ref{fig:complexG} is homologically zero.\footnote{The difference consists of three vertical contours (red, green and blue), with the green having twice the weight and opposite sign compared to the red and the blue.}
So, in some sense, one could say that the small black hole saddle ``does contribute'': Pick one or the other sign of $\text{Im}(G)$ and the result will contain a contribution from all three saddles.

However, in the limit of vanishing $\text{Im}(G)$, the green contour gives a purely imaginary contribution that only serves to cancel the imaginary contributions coming from the red and blue contours.
So, in some sense, the contribution from the green contour is fake.
The way to cleanly cancel the imaginary parts is to take a homology average of the contours shown in the two panels in figure \ref{fig:complexG}.\footnote{This procedure of averaging is known as median resummation in the context of Borel--\'Ecalle resummation. See, for example, Appendix A of \cite{Marino:2021lne}.} 
This is guaranteed to give the correct result since we argued that the two answers are the same.
However, what we gain by doing the average homology sum is that the two green contours cancel each other completely.
So, in this way of looking at the problem, the small black hole does not contribute to $Z(\beta)$.

To summarize this brief note, we posed and resolved a puzzle in the low-temperature confined phase of holographic field theories.
AdS black holes continue to exist as complex saddle points at low temperatures and, despite having an on-shell action whose real part is smaller than that of thermal AdS, they do not contribute to the thermal partition function.
We also discussed the question of whether the unstable small black hole contributes to the partition function in the high-temperature phase.
An extension of our analysis to the physically interesting case of rotating black holes is underway \cite{Singhi_to_appear}.
The simple mini-superspace perspective presented here may help illuminate aspects of more intricate problems, such as the case with a finite cutoff radius \cite{DiTucci:2020weq,Banihashemi:2025qqi}.

\paragraph{Acknowledgments.}
We would like to thank Yiming Chen, Edgar Shaghoulian, Arvin Shahbazi-Moghaddam and D. Stanford for discussions on related topics and Edgar Shaghoulian for comments on the draft.
We acknowledge support of the Department of Atomic Energy, Government of India, under project no. RTI4001.

\bibliographystyle{apsrev4-1long}
\bibliography{main}
\end{document}